\documentclass[journal]{assets/ieeeaccess}

\usepackage{cite}
\usepackage[flushleft]{threeparttable} 
\usepackage{amsmath, amssymb, amsfonts}
\usepackage{graphicx}
\usepackage{multirow} 
\usepackage{url} 
\usepackage{soul} 
\soulregister\cite7 
\soulregister\ref7 
\usepackage{tikz} 

\NewSpotColorSpace{PANTONE}
\AddSpotColor{PANTONE} {PANTONE3015C} {PANTONE\SpotSpace 3015\SpotSpace C} {1 0.3 0 0.2}
\SetPageColorSpace{PANTONE}%

\def\BibTeX{{\rm B\kern-.05em{\sc i\kern-.025em b}\kern-.08em
  T\kern-.1667em\lower.7ex\hbox{E}\kern-.125emX}}
 \def\code#1{\texttt{#1}}

\newcommand{\calc}[2]{\pgfmathparse{#1}\pgfmathprintnumber[assume math mode=true,fixed zerofill,precision=#2]{\pgfmathresult}}

\newcommand{\nPartDis}{100}
\newcommand{\nPartSha}{102}
\newcommand{\nPartTotal}{202}
\newcommand{\nPartUnique}{180}
\newcommand{\nPartBoth}{22}
\newcommand{\nPartTrainDis}{61}
\newcommand{\nPartValidDis}{20}
\newcommand{\nPartTestDis}{19}
\newcommand{\nPartTrainSha}{63}
\newcommand{\nPartValidSha}{20}
\newcommand{\nPartTestSha}{19}

\newcommand{\meanPartTotalDis}{816.07}
\newcommand{\meanPartTrainDis}{804.98}
\newcommand{\meanPartValidDis}{793.14}
\newcommand{\meanPartTestDis}{875.77}
\newcommand{\meanPartTotalSha}{830.61}
\newcommand{\meanPartTrainSha}{838.55}
\newcommand{\meanPartValidSha}{811.31}
\newcommand{\meanPartTestSha}{824.60}

\newcommand{\sdPartTotalDis}{237.47}
\newcommand{\sdPartTrainDis}{238.64}
\newcommand{\sdPartValidDis}{254.36}
\newcommand{\sdPartTestDis}{217.40}
\newcommand{\sdPartTotalSha}{240.80}
\newcommand{\sdPartTrainSha}{259.76}
\newcommand{\sdPartValidSha}{221.04}
\newcommand{\sdPartTestSha}{201.68}

\newcommand{\nIntakeTotalDis}{4790}
\newcommand{\nIntakeTrainDis}{2907}
\newcommand{\nIntakeValidDis}{943}
\newcommand{\nIntakeTestDis}{940}
\newcommand{\nIntakeTotalSha}{4279}
\newcommand{\nIntakeTrainSha}{2574}
\newcommand{\nIntakeValidSha}{896}
\newcommand{\nIntakeTestSha}{809}
\newcommand{\nServeTotalSha}{556}
\newcommand{\nServeTrainSha}{337}
\newcommand{\nServeValidSha}{107}
\newcommand{\nServeTestSha}{112}

\newcommand{\nIntakeTotal}{9069}

\newcommand{\meanIntakeTotalDis}{2.32}
\newcommand{\meanIntakeTrainDis}{2.36}
\newcommand{\meanIntakeValidDis}{2.24}
\newcommand{\meanIntakeTestDis}{2.29}
\newcommand{\meanIntakeTotalSha}{2.36}
\newcommand{\meanIntakeTrainSha}{2.44}
\newcommand{\meanIntakeValidSha}{2.23}
\newcommand{\meanIntakeTestSha}{2.24}
\newcommand{\meanServeTotalSha}{9.92}
\newcommand{\meanServeTrainSha}{9.87}
\newcommand{\meanServeValidSha}{10.99}
\newcommand{\meanServeTestSha}{9.05}

\newcommand{\sdIntakeTotalDis}{1.02}
\newcommand{\sdIntakeTrainDis}{1.04}
\newcommand{\sdIntakeValidDis}{0.98}
\newcommand{\sdIntakeTestDis}{1.01}
\newcommand{\sdIntakeTotalSha}{1.14}
\newcommand{\sdIntakeTrainSha}{1.16}
\newcommand{\sdIntakeValidSha}{1.17}
\newcommand{\sdIntakeTestSha}{1.02}
\newcommand{\sdServeTotalSha}{4.97}
\newcommand{\sdServeTrainSha}{4.52}
\newcommand{\sdServeValidSha}{6.56}
\newcommand{\sdServeTestSha}{4.34}

\begin{document}
\history{Date of publication xxxx 00, 0000, date of current version xxxx 00, 0000.}
\doi{10.1109/ACCESS.2020.3026965}

\title{OREBA: A Dataset for Objectively Recognizing Eating Behaviour and Associated Intake}
\author{\uppercase{Philipp V. Rouast}\authorrefmark{1}, \IEEEmembership{Student~Member, IEEE}, \uppercase{Hamid Heydarian}\authorrefmark{1}, \uppercase{Marc T. P. Adam}\authorrefmark{1,3}, and \uppercase{Megan E. Rollo}\authorrefmark{2,3}}%
\address[1]{School of Electrical Engineering and Computing, The University of Newcastle, Callaghan, NSW 2308, Australia}%
\address[2]{School of Health Sciences, The University of Newcastle, Callaghan, NSW 2308, Australia}
\address[3]{Priority Research Centre for Physical Activity and Nutrition, The University of Newcastle, Callaghan, NSW 2308, Australia}

\tfootnote{We gratefully acknowledge the support by the Bill \& Melinda Gates Foundation [OPP1171389]. Philipp Rouast and Hamid Heydarian were supported by an Australian Government Research Training (RTP) Scholarship.}

\pubid{ \begin{minipage}{\textwidth}\centering\tiny
This article has been accepted for publication in a future issue of this journal, but has not been fully edited.\\Content may change prior to final publication. Citation information: DOI 10.1109/ACCESS.2020.3026965 
\end{minipage}}

\markboth{Author \headeretal: Preparation of Papers for IEEE TRANSACTIONS and JOURNALS}{Author \headeretal: Preparation of Papers for IEEE TRANSACTIONS and JOURNALS}

\corresp{Corresponding author: Marc T. P. Adam (e-mail: marc.adam@newcastle.edu.au).}

\begin{abstract}
Automatic detection of intake gestures is a key element of automatic dietary monitoring.
Several types of sensors, including inertial measurement units (IMU) and video cameras, have been used for this purpose. 
The common machine learning approaches make use of the labeled sensor data to automatically learn how to make detections.
One characteristic, especially for deep learning models, is the need for large datasets.
To meet this need, we collected the Objectively Recognizing Eating Behavior and Associated Intake (OREBA) dataset.
The OREBA dataset aims to provide comprehensive multi-sensor data recorded during the course of communal meals for researchers interested in intake gesture detection.
Two scenarios are included, with {\nPartDis} participants for a discrete dish and {\nPartSha} participants for a shared dish, totalling {\nIntakeTotal} intake gestures.
Available sensor data consists of synchronized frontal video and IMU with accelerometer and gyroscope for both hands.
We report the details of data collection and annotation, as well as details of sensor processing.
The results of studies on IMU and video data involving deep learning models are reported to provide a baseline for future research.
Specifically, the best baseline models achieve performances of $F_1=0.853$ for the discrete dish using video and $F_1=0.852$ for the shared dish using inertial data.
\end{abstract}


\titlepgskip=-15pt

\maketitle


\section{Introduction}
\label{sec:introduction}

\IEEEPARstart{T}{raditional} dietary assessment methods are reliant on self-report data.
While data captured with active methods such as self-report and 24-hr recall are widely used in practice, they are not without limitations (e.g., human error, time-consuming manual process) \cite{lichtman1992discrepancy}.
Automatic dietary monitoring, where data is collected and processed independent of the individual, has the potential to complement data from traditional methods and reduce associated biases \cite{rouast2018using}.
In addition, such systems have the potential to support personal self-monitoring solutions by providing individuals with targeted eating behaviour recommendations. 

A key element of automatic dietary monitoring is the detection of \textit{intake gestures} (i.e., the process of moving food or drink towards the mouth).
Recent research on this task focuses mainly on machine learning approaches which are characterized by a need for large amounts of labeled data.
This is especially true in conjunction with deep learning, which has been applied in this context since 2017 \cite{kyritsis2017food}.
However, collecting, synchronizing, and labeling data of eating occasions is a work-intensive process.
Hence, there is a need for more public datasets to reduce barriers for researchers to create new machine learning models, and to objectively compare the performance of existing approaches \cite{qiu2019assessing}, \cite{konstantinidis2020validation}. 

At the same time, current research on dietary monitoring identified a gap in research on shared plate eating \cite{burrows2019dietary}.
Communal eating (i.e., eating occasions involving more than one person) is not yet well understood, let alone the impact it has on accuracy of automatic dietary monitoring.
Hence, existing research on capturing dietary intake from discrete dishes needs to be complemented and contrasted with research on the detection of intake from shared dishes.

\makeatletter
\newcommand{\thickhline}{%
  \noalign {\ifnum 0=`}\fi \hrule height 1pt
  \futurelet \reserved@a \@xhline
}

\begin{table*}[ht]
\centering
\caption{Public datasets of intake gestures with synchronized sensor data and annotations available.}
\label{tab:datasets}
\begin{threeparttable}
\begin{tabular}{ l | c c c c | c | c c | c }
& \multicolumn{4}{c|}{Synchronized sensors} & Ground & Participants & Intake & \\
Dataset & Video & Audio & IMU & Scale & truth & (Recordings)\tnote{a} & events & Annotations \\
\thickhline
\multirow{2}{*}{ACE \cite{merck2016multimodality}} & \multirow{2}{*}{-} & \multirow{2}{*}{Earbud} & Both hands & \multirow{2}{*}{-} & Video & \multirow{2}{*}{7 (13)} & \multirow{2}{*}{1492} & Chews, swallows with type and amount \\
 & & & 15 Hz & & multi\tnote{b} & & & of food and drink. \\
\hline
\multirow{2}{*}{Clemson \cite{shen2017assessing}} & \multirow{2}{*}{-} & \multirow{2}{*}{-} & Dominant hand & Tray & Video & \multirow{2}{*}{264 (488)} & \multirow{2}{*}{20644} & Intake gestures, utensiling. Intake annotated \\
 & & & 15 Hz & 15 Hz & ceiling\tnote{b} & & & with hand, utensil, container, and food. \\
\hline
\multirow{2}{*}{FIC \cite{kyritsis2019modeling}} & \multirow{2}{*}{-} & \multirow{2}{*}{-} & Dominant hand & \multirow{2}{*}{-} & Video & \multirow{2}{*}{12 (21)} & \multirow{2}{*}{1332} & Plain intake gestures; further \\
 & & & 100 Hz & & frontal\tnote{b} & & & annotation of micromovements. \\
\hline
\hline
\multirow{2}{*}{OREBA-DIS} & Frontal & \multirow{2}{*}{-} & Both hands & \multirow{2}{*}{-} & Video & \multirow{2}{*}{{\nPartDis} ({\nPartDis})} & \multirow{2}{*}{{\nIntakeTotalDis}} & Intake gestures annotated with \\
 & 24 fps & & 64 Hz & & frontal & & & eat/drink, hand and utensil. \\
\hline
\multirow{2}{*}{OREBA-SHA} & Frontal & \multirow{2}{*}{-} & Both hands & Communal & Video & \multirow{2}{*}{{\nPartSha} ({\nPartSha})} & \multirow{2}{*}{{\nIntakeTotalSha}} & Intake gestures annotated with \\
 & 30 fps & & 64 Hz & 1 Hz & multi & & & eat/drink, hand and utensil. \\
\thickhline
\end{tabular}
\begin{tablenotes}
\item[a] One recordings equals one person consuming one meal.
\item[b] Used for ground truth annotation, but not available for download as of early 2020.
\end{tablenotes}
\end{threeparttable}
\end{table*}

In order to address these gaps, the present paper introduces the \underline{O}bjectively \underline{R}ecognizing \underline{E}ating \underline{B}ehavior and \underline{A}ssociated Intake (OREBA) dataset.
The goal of OREBA is to facilitate the automatic detection of intake gestures in communal eating across two scenarios (discrete dish and shared dish).
By creating this dataset and making it available to the wider research community, this paper makes four key contributions:

\begin{enumerate}
	\item \textbf{Large-scale dataset:} We conducted a total of {\nPartTotal} meal recordings, with {\nPartUnique} unique individuals participating who consented to their data being used by other research institutions. In total, we captured {\nIntakeTotal} intake gestures from discrete and shared dishes. Two independent annotators labeled and cross-checked the intake gestures.
	\item \textbf{Public availability:} Progress in the research of machine learning methods is tightly linked to the public availability of labeled datasets (e.g. \cite{kay2017kinetics}, \cite{deng2009imagenet}, \cite{mollahosseini2017affectnet} the dataset publicly available to researchers, OREBA can be used to objectively benchmark existing and emerging machine learning approaches. At the same time, it reduces the time-consuming burden for researchers to collect, annotate and cross-check their own data.
	\item \textbf{Communal eating:} While existing research has provided important insights into automatically detecting human intake gestures in individual settings, research on communal eating is scant \cite{heydarian2019assessing} \cite{burrows2019dietary}. To the best of our knowledge, this is the first dataset capturing communal eating from both discrete and shared dishes.
	\item \textbf{Multiple modalities:} The dataset includes synchronized frontal video and inertial measurement unit (IMU) sensor data from both hands, along with labels for each intake gesture.  A single spherical camera positioned in the center of the table made it possible to capture the entire communal eating scene of up to four participants, offering a full view of all relevant gestures. While existing inertial datasets on intake gesture detection often use video as ground truth, none of the existing datasets currently include video data as part of the synchronized sensor data for analysis.
\end{enumerate}

In the following, Section \ref{sec:relatedwork} gives an overview of the related work and existing datasets, Section \ref{sec:oreba} introduces the data collection and annotation process of the OREBA dataset in detail, Section \ref{sec:baseline} provides results from our initial studies as baselines.
Finally, we provide a discussion in Section \ref{sec:discussion} and conclusions in Section \ref{sec:conclusions}.

\section{Related work}
\label{sec:relatedwork}

\Figure[ht]()[width=.96\textwidth]{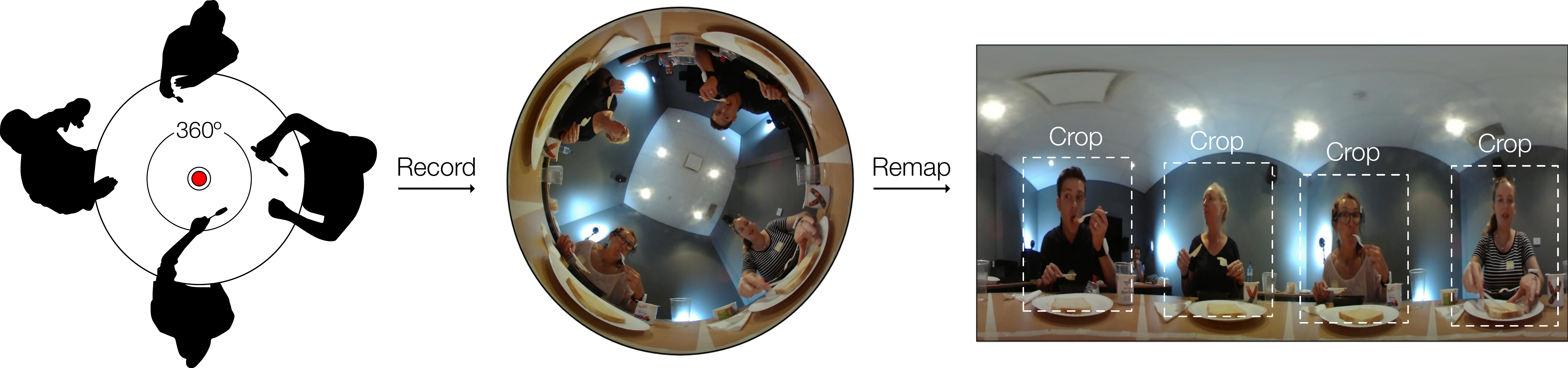}
{The spherical video is remapped to equirectangular representation, cropped, and reshaped to square shape.\label{fig:recording}}

\Figure[ht]()[width=.96\columnwidth]{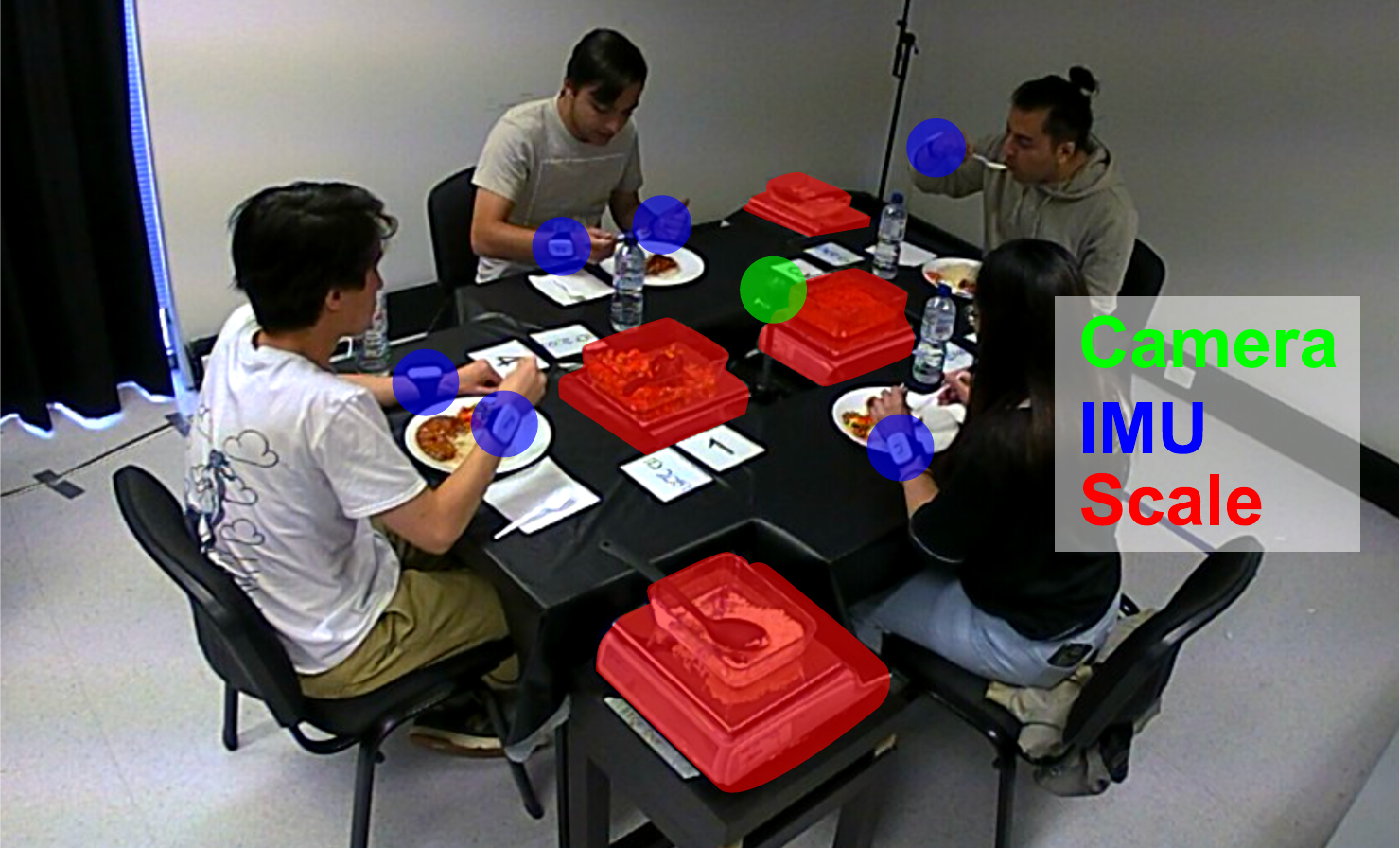}
{Study setup for OREBA-SHA. One camera in the center of the table, IMU on each wrist, and four scales.\label{fig:setup}}

\subsection{Automatic dietary monitoring}

Automatic dietary monitoring encompasses three major goals: (i) detecting the timing of intake events, (ii) recognizing the type of food or drink, and (iii) estimating the weight consumed.
Detection of intake behavior, which is associated with intake gestures, chews, and swallows, can be considered as part of the first goal.
Researchers have leveraged various different sensor types for this purpose.
While chews and swallows can be detected using audio signals \cite{zhang2018estimating}, intake gestures are typically handled using an IMU including accelerometer and gyroscope sensors \cite{zhang2018sense} \cite{heydarian2019assessing}.
Before the application of deep learning architectures, the traditional approach in this field reduced the dimensionality of the raw sensor data by extracting handcrafted features based on expert knowledge. 
Deep learning methods have been explored to detect individual intake gestures with inertial sensor data since 2017 \cite{kyritsis2017food} and with video data since 2018 \cite{rouast2018using}, \cite{qiu2019assessing}, \cite{konstantinidis2020validation}, whereby large amounts of labeled examples are leveraged to let algorithms learn the features automatically.
The most widely used approach in this space builds on convolutional neural networks (CNN) and long short-term memory (LSTM) models \cite{kyritsis2020data}, however gated recurrent unit (GRU) models have also been applied, especially in the context of activity recognition in daily living \cite{zhu2018sequence}.

\subsection{Existing datasets}

To date, most published studies on recognition of intake behaviour rely on dedicated, private datasets collected for a specific purpose.
Considering the shift towards adoption of deep learning techniques, we expect an increasing need for large, public datasets that existing and emerging machine learning approaches can objectively be benchmarked on.
Similar developments can be observed across several related fields such as action recognition \cite{kay2017kinetics}, affect recognition \cite{rouast2019deep} \cite{mollahosseini2017affectnet}, and object recognition \cite{deng2009imagenet}.

Table \ref{tab:datasets} provides an overview of publicly available datasets on intake gestures which feature synchronized sensor data of eating occasions with labels for individual gestures (intake gestures or other eating related gestures)\footnote{A related dataset is iHEARu-EAT \cite{hantke2016hear}, however we did not include it here since it does not focus on intake events.}.

\begin{itemize}
	\item The \textit{accelerometer and audio-based calorie estimation (ACE) dataset}\footnote{See \url{http://www.skleinberg.org/data.html}} \cite{merck2016multimodality} contains seven participants with audio and IMU data for both hands and the head. Annotations of type and amount of food and drink are available for chews and swallows.
	\item The \textit{Clemson Cafeteria dataset}\footnote{See \url{http://cecas.clemson.edu/~ahoover/cafeteria/}. Recordings with missing annotations are excluded here.} \cite{shen2017assessing} contains 264 participants and 488 recordings. IMU data is available at 15 Hz for the dominant hand, along with scale measurements for the tray. Each intake gesture is annotated with hand, utensil, container, and food.
	\item The \textit{Food Intake Cycle (FIC) dataset}\footnote{See \url{https://mug.ee.auth.gr/intake-cycle-detection/}} \cite{kyritsis2019modeling}, which consists of 12 participants and 21 recordings, includes IMU data for the dominant hand. The focus is on the micromovements during intake gestures.
\end{itemize}

While video is commonly used as ground truth, none of the existing datasets currently include video data as part of the synchronized sensor data for analysis.
In terms of IMU data and quantity of recorded intake events, we find that the existing datasets are restricted either to data from only one hand, a relatively low recording frequency (15 Hz), or few participants.
We aim to further the field by establishing the OREBA dataset, which includes video and IMU from both hands, at a quantity of intake events sufficient to train deep learning models for both video and inertial modalities.
 
\section{The OREBA dataset}
\label{sec:oreba}

The OREBA dataset aims to provide a comprehensive multi-sensor recording of communal intake occasions for researchers interested in automatic detection of intake gestures and other behaviours associated with intake  (e.g., serving food onto a plate).
Available sensor data consists of synchronized frontal video and accelerometer and gyroscope for both hands in two different scenarios (i.e., discrete dish and shared dish).
IRB approval was given (H-2017- 0208), and the data was recorded between Mar 2018 and Oct 2019.

\subsection{Scenarios}

The OREBA dataset consists of two separate communal eating scenarios. In each scenario, groups of up to four participants were simultaneously recorded consuming a meal at a communal table:

\begin{enumerate}
	\item OREBA-DIS: In the first scenario, foods were served in discrete portions to each participant. The meal consisted of lasagna (choice between vegetarian and beef), bread, and yogurt. Additionally, there was water available to drink, and butter to spread on the bread. The study setup for OREBA-DIS is shown in Fig. \ref{fig:recording}.
	\item OREBA-SHA: In the second scenario, participants consumed a communal dish of vegetable korma or butter chicken with rice and mixed vegetables. Additionally, there was water available to drink. The study setup for OREBA-SHA is shown in Fig. \ref{fig:setup}.
\end{enumerate}

Lasagna and rice-based dishes were chosen since they are amongst the most common dishes in similar studies \cite{heydarian2019assessing}.
All participants in each scenario are unique, however {\nPartBoth} participants participated in both scenarios.

\subsection{Sensors}

For each group, video was recorded using a spherical camera placed in the center of the shared table (360fly-4K\footnote{See \url{https://www.360fly.com/}}).
This allowed video recording to occur in a simultaneous and unobtrusive way for all participants engaging in the communal eating occasion around the table.
The sampling rates are 24 fps for OREBA-DIS, and 30 fps for OREBA-SHA.
Each participant wore two IMUs, one on each wrist (Movisens Move 3+ \footnote{See \url{https://www.movisens.com/en/products/activity-sensor-move-3/}}).
The IMU included an accelerometer and a gyroscope with a sampling rate of 64 Hz.
For OREBA-SHA, four scales additionally recorded the weight of the communal dishes (two rices dishes, one wet dish, one vegetable dish) at 1 Hz (Adam Equipment CBK 4).

\subsection{Sensor processing}

\subsubsection{Video}

As shown in Fig. \ref{fig:recording}, we first mapped the spherical video from the 360-degree camera to equirectangular representation\footnote{See \url{https://github.com/prouast/equirectangular-remap}}.
Then, we separated the equirectangular representation into individual participant videos by cropping the areas of interest.
We further resized each participant video to a square shape.
The two spatial resolutions 140x140 (e.g., \code{<id>\_video\_140p.mp4}) and 250x250 pixels (e.g., \code{<id>\_video\_250p.mp4}) are included.
All videos are encoded using the H.264 standard and stored in mp4 containers.

\subsubsection{Inertial Measurement Unit}

Raw accelerometer data is measured in $g$, while gyroscope data is measured in $deg/s$.
The OREBA dataset includes (i) raw sensor data without any processing for left and right hand (e.g., \code{<id>\_inertial\_raw.csv}), and (ii) processed sensor data for dominant and non-dominant eating hand (e.g., \code{<id>\_inertial\_processed.csv}).
Raw data is included since a recent study on OREBA indicates that data preprocessing only marginally improves results when combined with deep learning \cite{heydarian2020deep}.
Processed data is generated from the raw data according to the following steps:

\textit{1: Removal of gravity effect.}
The raw accelerometer reading is subject to acceleration from participants' wrist movements as well as the earth's gravitational field.
We remove this gravity effect by estimating sensor orientation using sensor fusion with Madgwick's filter \cite{madgwick2010efficient}, rotation of the acceleration vector with the resulting quaternion, and deduction of the gravity vector (see \cite{kyritsis2017food} for a similar approach).

\textit{2: Standardization.}
Each column (i.e. each axis for each modality and hand) is standardized by subtracting its mean and dividing by its standard deviation (see \cite{kyritsis2019modeling} for a similar approach).
Processed data can hence be regarded as unitless.

\begin{figure}[ht]
\centering
\includegraphics[width=\columnwidth]{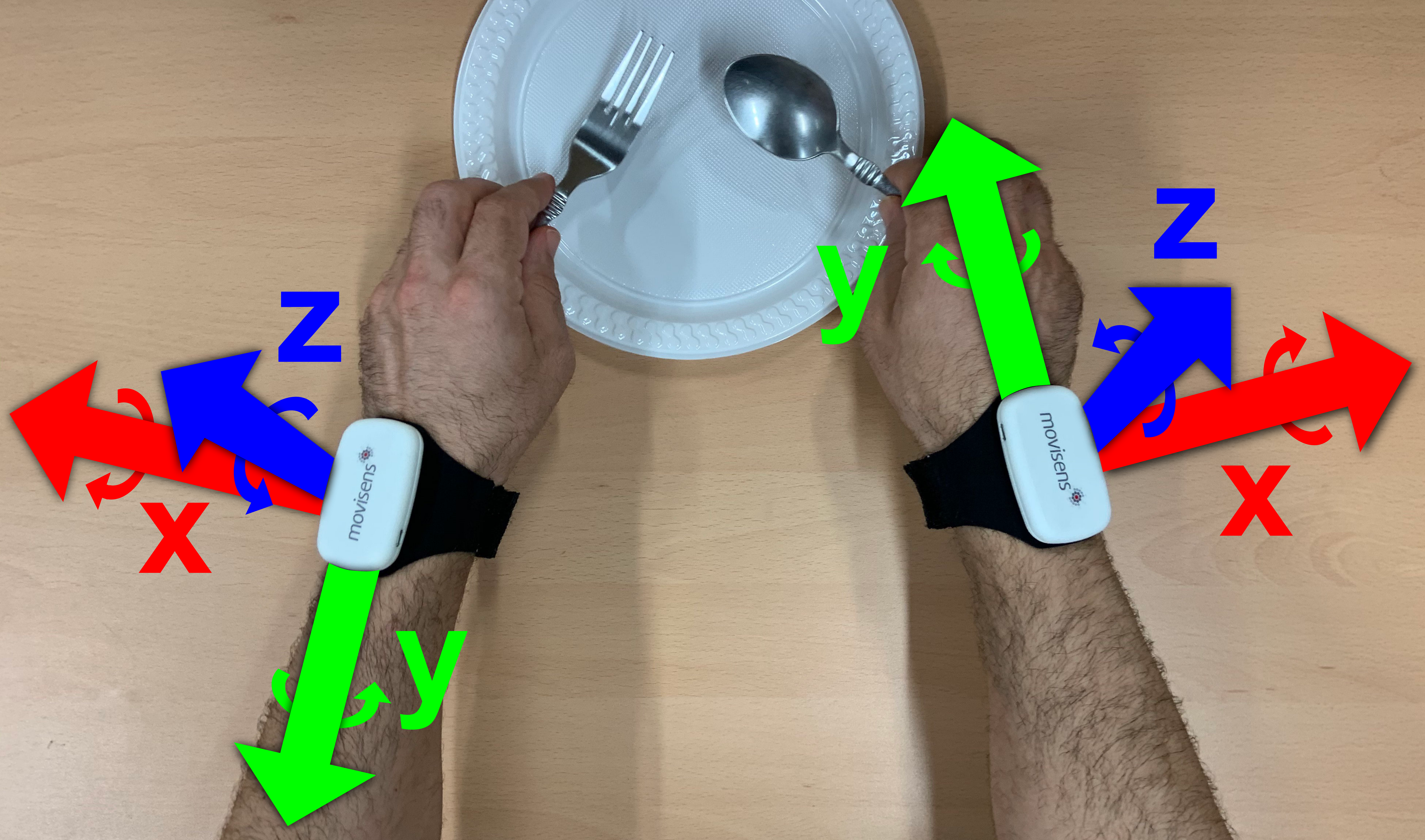}
\caption{The wrist-worn sensors with their internal coordinate frames. Additionally, the direction of positive rotations are indicated for each axis.}
\label{fig:sensors}
\end{figure}

\textit{3: Transforming from left and right hand to dominant and non-dominant hand.}
To achieve data uniformity, we report hands in the processed data as \textit{dominant} and \textit{non-dominant}.
A similar approach was chosen for the FIC dataset \cite{kyritsis2020data}.
All data reported as dominant hands correspond to right hands, and non-dominant hands to left hands; for left-handed participants data for both hands has been transformed to achieve this.
Specifically, we mirrored the data of left-handed participants to transform the data from the left wrist as if it has been recorded on the right wrist, and vice versa.
Due to the way the sensors are mounted on the wrist (see Fig. \ref{fig:sensors}), the horizontal direction corresponds to the x axis.
For accelerometer data, we estimate mirroring by flipping the sign of the x axis, and for gyroscope by flipping the signs of the y and z axis.
Further, we also flip the signs of the x and y axis to compensate for the different sensor orientations on the wrists, yielding transformation (\ref{eq:acc}) for accelerometer and (\ref{eq:gyro}) for gyroscope.

\setlength{\abovedisplayskip}{0pt} \setlength{\abovedisplayshortskip}{0pt}

\begin{equation}
	[a'_x,a'_y,a'_z] = [-(-a_x),-a_y,a_z] = [a_x,-a_y,a_z] \label{eq:acc}
\end{equation}

\begin{equation}
	[g'_x,g'_y,g'_z] = [-g_x,-(-g_y),-g_z] = [-g_x,g_y,-g_z] \label{eq:gyro}
\end{equation}

Note that the mirroring technique propopsed here could also be of use for data augmentation pipelines.

\subsubsection{Synchronization}

Ground truth for sensor synchronization was acquired by asking participants to clap their hands before starting, and after finishing their meal (see \cite{kyritsis2017food} for a similar approach).
The clapping creates a distinct signature in both the video recording and the accelerometer.
All sensors were trimmed in time and synchronized for each participant based on these two reference points.

\subsubsection{Scales}

In addition to detecting individual intake gestures, there is also a growing body of research on determining the amounts of food consumed based on continuous weight measurement using scales \cite{shen2017assessing} \cite{papapanagiotou2019automatic}.
By detecting changes in the amounts of food on a plate, the scale data can complement other modalities in detecting intake and/or serving gestures as well as evaluating the amounts of food consumed from specific plates.
The shared plate setting in OREBA-SHA included four scales that measured the weight of the two rice dishes at two corners of the table as well as the wet dish and the vegetable dish in the centre of the table (see Figure \ref{fig:setup}).
These scales recorded the weight of the four dishes in grams at a sampling rate of 1 Hz.
The scale recordings were time-synchronized by means of a time-lapse camera and a 200g calibration weight.
At the start of the recording, a research assistant removed the calibration weight from the scale.
This was captured by the scale recordings as well as the time-lapse camera.
Further, the time-lapse camera also captured the clapping at the start of the recording.
Based on this, each scale recording includes a reference in seconds to the clapping at the start of a recording.
Further, the dataset provides a mapping of each participant number to the closest rice dish.

\subsection{Annotation}

\begin{table}[t]
\centering
\caption{The labeling scheme.}
\label{tab:labeling}
\begin{tabular}{ l | l }
Category & Possible values \\ 
\thickhline
Main & Intake, Serve \\
\hline
Sub & Intake-Eat, Intake-Drink, \\
& Serve-Self, Serve-Other \\
\hline
Hand & Left, Right, Both \\
\hline
Utensil & Fork, Spoon, Hand, Knife, Finger, Cup, Bottle \\
\thickhline
\end{tabular}
\end{table}

\begin{figure*}[ht]
\centering
\includegraphics[width=.95\textwidth]{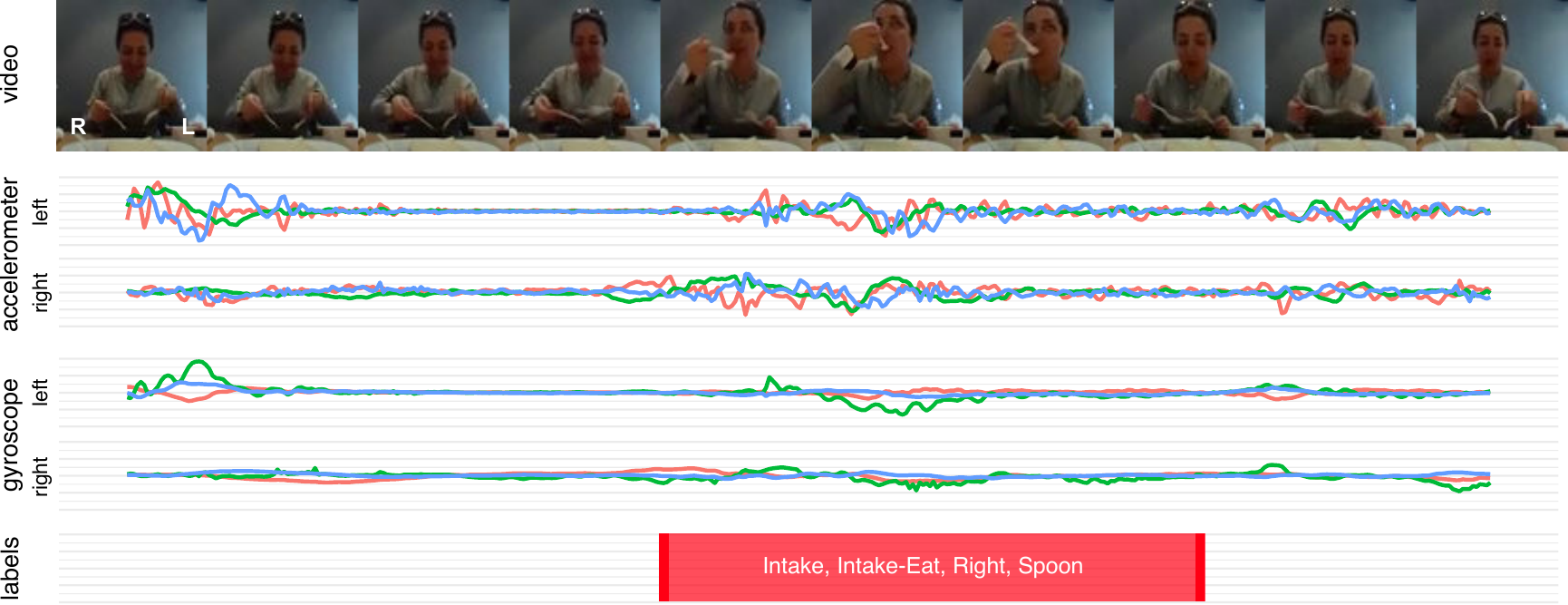}
\caption{Example of a labeled intake gesture with video, accelerometer, and gyroscope sensor data. For easier display, the video framerate has been reduced.}
\label{fig:example}
\end{figure*}

All relevant gestures were labeled and cross-checked by two independent annotators using ChronoViz \footnote{See \url{http://chronoviz.com}}.
Each gesture includes a start and an end timestamp:

\begin{itemize}
	\item The start timestamp is the point where the final uninterrupted movement to execute the gesture starts;
	\item the end timestamp is the point when the participant has finished returning their hand(s) from the movement or started a different gesture.
\end{itemize}

Additionally, each gesture is assigned four labels according to our labeling scheme as listed in Table \ref{tab:labeling}.
Besides the \textit{Main} identification as an Intake or Serve gesture, this scheme allows to further specify a \textit{Sub} category for each gesture, as well as indicating the \textit{Hand} and \textit{Utensil} (e.g., \code{<id>\_annotations.csv}).

The scheme is designed to be extendable with more categories in possible extensions of the dataset.
The discrete dish scenario OREBA-DIS includes Intake labels.
Correspondingly, the shared dish scenario OREBA-SHA includes both Intake and Serve labels.
Figure \ref{fig:example} depicts an example of a labeled intake gesture and associated IMU sensor data.

\subsection{Splits}
\label{sec:oreba:sub:splits}

For machine learning problems with time-intensive training and evaluation, the best practice is to train, validate, and test using three separate sets of data \cite{goodfellow2016deep}.
Models are trained with the training set, hyperparameters are tuned using the validation set, and reported results are based on the test set.
We choose a split of approximately 3:1:1, such that each participant only appears in one of the three subsets; this is to ensure that we are measuring the model's ability to generalise and avoid data leakage.
The recommended split is included in the dataset download.
Table \ref{tab:oreba-summary} summarises high-level statistics on these splits.

\begin{table*}[t]
\centering
\caption{Summary statistics for our dataset and the training/validation/test split.}
\label{tab:oreba-summary}
\begin{tabular}{l l | c c c | c c c | c c c | c c c }
& & \multicolumn{3}{c|}{Training} & \multicolumn{3}{c|}{Validation} & \multicolumn{3}{c|}{Test} & \multicolumn{3}{c}{Total} \\
Scenario & Type & \# & Mean [s] & Std [s] & \# & Mean [s] & Std [s] & \# & Mean [s] & Std [s] & \# & Mean [s] & Std [s] \\
\thickhline
\multirow{2}{*}{OREBA-DIS} & Participants & {\nPartTrainDis} & {\meanPartTrainDis} & {\sdPartTrainDis} & {\nPartValidDis} & {\meanPartValidDis} & {\sdPartValidDis} & {\nPartTestDis} & {\meanPartTestDis} & {\sdPartTestDis} & {\nPartDis} & {\meanPartTotalDis} & {\sdPartTotalDis} \\
& Intake Gest. & {\nIntakeTrainDis} & {\meanIntakeTrainDis} & {\sdIntakeTrainDis} & {\nIntakeValidDis} & {\meanIntakeValidDis} & {\sdIntakeValidDis} & {\nIntakeTestDis} & {\meanIntakeTestDis} & {\sdIntakeTestDis} & {\nIntakeTotalDis} & {\meanIntakeTotalDis} & {\sdIntakeTotalDis} \\
\hline
\multirow{3}{*}{OREBA-SHA} & Participants & {\nPartTrainSha} & {\meanPartTrainSha} & {\sdPartTrainSha} & {\nPartValidSha} & {\meanPartValidSha} & {\sdPartValidSha} & {\nPartTestSha} & {\meanPartTestSha} & {\sdPartTestSha} & {\nPartSha} & {\meanPartTotalSha} & {\sdPartTotalSha} \\
& Intake Gest. & {\nIntakeTrainSha} & {\meanIntakeTrainSha} & {\sdIntakeTrainSha} & {\nIntakeValidSha} & {\meanIntakeValidSha} & {\sdIntakeValidSha} & {\nIntakeTestSha} & {\meanIntakeTestSha} & {\sdIntakeTestSha} & {\nIntakeTotalSha} & {\meanIntakeTotalSha} & {\sdIntakeTotalSha} \\
& Serve Gest. & {\nServeTrainSha} & {\meanServeTrainSha} & {\sdServeTrainSha} & {\nServeValidSha} & {\meanServeValidSha} & {\sdServeValidSha} & {\nServeTestSha} & {\meanServeTestSha} & {\sdServeTestSha} & {\nServeTotalSha} & {\meanServeTotalSha} & {\sdServeTotalSha} \\
\thickhline
\end{tabular}
\end{table*}

\subsection{Demographics}

Out of {\nPartUnique} participants in total, 161 agreed to complete a demographics questionnaire.
Across the dataset, 67\% identified as male and 33\% as female.
The median age is 24, with the minimum and maximum age being 18 and 54 years respectively.
Reported ethnicities in the dataset include White Australian (52.2\%), White other European (9.9\%), Chinese (8.7\%), Other Asian (8.7\%), Persian (5.6\%), Arabic (3.1\%), White British (3.1\%), African (2.5\%), and South East Asian (1.8\%).  
About 10\% reported being left-, and 90\% right-handed.

\subsection{Availability}
\label{sec:oreba:sub:availability}

The OREBA dataset is available on request to research groups at academic institutions.
Please visit \url{http://www.newcastle.edu.au/oreba} to download the data sharing agreement and get access.

\section{Baseline for intake gesture detection}
\label{sec:baseline}

Intake gesture detection refers to the task of detecting the times of individual intake gestures from sensor data.
Similar to dataset papers in other areas \cite{deng2009imagenet} \cite{mollahosseini2017affectnet}, we provide baseline results for this task on OREBA-DIS and OREBA-SHA.
We apply the two-stage approach proposed by Kyritsis et al. \cite{kyritsis2019modeling} to estimate frame-level intake probabilities and detect intake gestures.
For this purpose, we train separate baseline models on inertial and video sensor data, introduced in Section \ref{sec:baseline:sub:models}.
To ensure comparability with future studies, we use the publicly available data splits introduced in Section \ref{sec:oreba:sub:splits} for training, validation, and test; we additionally report details on training and used evaluation metric in Section \ref{sec:baseline:sub:training}.

\subsection{Baseline models}
\label{sec:baseline:sub:models}

For each modality, we use one simple CNN and one more complex model proposed in previous studies \cite{heydarian2020deep} \cite{rouast2019learning}.
As listed in Table \ref{tab:baseline}, this results in a total of eight baseline models, considering the different scenarios, modalities, and models.

\subsubsection{Inertial}

The inertial models are taken from a recent study on OREBA by Heydarian et al. \cite{heydarian2020deep}.
We compare the simple CNN with the more complex CNN-LSTM proposed in the aforementioned work.
The simple CNN consists of seven CNN layers and one fully-connected layer, with one max pooling layer following each CNN layer.
The CNN-LSTM consists of four CNN layers with 128 kernels each, two LSTM layers with 64 units each, and one fully-connected layer.
Full details on these models are available in the Supplemental Material S1.

\subsubsection{Video}

The video models are taken from a recent study on OREBA by Rouast et al. \cite{rouast2019learning}.
For our comparison we use the simple CNN and the more complex ResNet-50 SlowFast proposed by Rouast et al.
While the simple CNN only uses one frame at a time, the ResNet-50 SlowFast model uses 16 frames.
The ResNet-50 SlowFast model consists of two 50-layer 3D CNNs which are fused with lateral connections and spatially aligned 2D conv fusion.
Full details on these models are available in the Supplemental Material S2.

\subsection{Training and evaluation metrics}
\label{sec:baseline:sub:training}

\subsubsection{Training}

We train a total of eight baseline models.
All baseline models are trained using the \textit{Adam} optimizer with an exponentially decaying learning rate on the respective training dataset.
We use batch size 256 for inertial data, and batch sizes 8 (ResNet-50 SlowFast) / 64 (CNN) for video data.
Model selection is done using the validation set.

\subsubsection{Evaluation metrics}

\begin{figure}[t]
\centering
\includegraphics[width=.8\columnwidth]{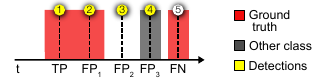}
\caption{The evaluation scheme (proposed by \cite{kyritsis2019modeling}; figure from \cite{rouast2019learning} extended here). (1) A true positive is the first detection within each ground truth event; (2) False positives of type 1 are further detections within the same ground truth event; (3) False positives of type 2 are detections outside ground truth events; (4) False positives of type 3 are detections made for the wrong class; (5) False negatives are non-detected ground truth events.}
\label{fig:scheme}
\end{figure}

\newcommand\tp{\mathit{TP}}
\newcommand\fn{\mathit{FN}}
\newcommand\fp{\mathit{FP}}

We extend the evaluation scheme proposed by Kyritsis et al. \cite{kyritsis2019modeling} as depicted in Fig. \ref{fig:scheme}.
The scheme uses the ground truth to translate sparse detections into measurable metrics for a given label category.
As Rouast and Adam \cite{rouast2019learning} report, one correct detection per ground truth event counts as a true positive ($\tp$), while further detections within the same ground truth event are false positives of type 1 ($\fp_1$).
Detections outside ground truth events are false positives of type 2 ($\fp_2$) and non-detected ground truth events count as false negatives ($\fn$).
The scheme has been extended here to support the multi-class case, where detections for a wrong class are false positives of type 3.
Based on the aggregate counts, precision ($\frac{\tp}{\tp+\fp_1+\fp_2+\fp_3}$), recall ($\frac{\tp}{\tp+\fn}$), and the $F_1$ score ($2*\frac{\mathit{Precision}*\mathit{Recall}}{\mathit{Precision}+\mathit{Recall}}$) can be calculated.

\subsection{Baseline results}


\newcommand{\fVideoBaseDis}{0.713}
\newcommand{\fVideoSOTADis}{0.853}

\newcommand{\fInertBaseDis}{0.758}
\newcommand{\fInertSOTADis}{0.778}


\newcommand{\fVideoBaseSha}{0.696}
\newcommand{\fVideoSOTASha}{0.808}

\newcommand{\fInertBaseSha}{0.839}
\newcommand{\fInertSOTASha}{0.852}

\DeclareRobustCommand{\relImpVideoDis}{\calc{(\fVideoSOTADis-\fVideoBaseDis)/\fVideoBaseDis*100}{1}}
\DeclareRobustCommand{\relImpVideoSha}{\calc{(\fVideoSOTASha-\fVideoBaseSha)/\fVideoBaseSha*100}{1}}

\DeclareRobustCommand{\relImpInertDis}{\calc{(\fInertSOTADis-\fInertBaseDis)/\fInertBaseDis*100}{1}}
\DeclareRobustCommand{\relImpInertSha}{\calc{(\fInertSOTASha-\fInertBaseSha)/\fInertBaseSha*100}{1}}

\begin{table*}[ht]
\centering
\caption{Baseline test set results for intake gesture detection. On OREBA-DIS, the video model performs better than the inertial model, while the opposite is true on OREBA-SHA. This indicates that the test set for OREBA-DIS is more challenging for using inertial data, while the test set for OREBA-SHA is more challenging for using video data.}
\label{tab:baseline}
\renewcommand{\arraystretch}{1.05}
\begin{threeparttable}
\begin{tabular}{l | l | l | c c c c | c }
Dataset & Modality & Method & $TP$ & $FP_1$ & $FP_2$ & $FN$ & $F_1$ \\
\thickhline
\multirow{4}{*}{OREBA-DIS} & \multirow{2}{*}{Video} & CNN \cite{rouast2019learning} & 668 & 28 & 242 & 267 & \fVideoBaseDis \\ 
& & ResNet-50 SlowFast \cite{rouast2019learning} & 749 & 20 & 52 & 186 & \textbf{\fVideoSOTADis} \\ 
\cline{2-8}
& \multirow{2}{*}{Inertial} & CNN \cite{heydarian2020deep} & 702 & 42 & 171 & 235 & \fInertBaseDis \\ 
& & CNN-LSTM \cite{heydarian2020deep} & 743 & 41 & 188 & 194 & \fInertSOTADis \\ 
\hline
\multirow{4}{*}{OREBA-SHA} & \multirow{2}{*}{Video} & CNN \cite{rouast2019learning} & 564 & 26 & 228 & 239 & \fVideoBaseSha \\ 
& & ResNet-50 SlowFast \cite{rouast2019learning} & 670 & 16 & 170 & 133 & \fVideoSOTASha \\ 
\cline{2-8}
& \multirow{2}{*}{Inertial} & CNN \cite{heydarian2020deep} & 719 & 46 & 146 & 84 & \fInertBaseSha \\ 
& & CNN-LSTM \cite{heydarian2020deep} & 732 & 34 & 149 & 71 & \textbf{\fInertSOTASha} \\ 
\thickhline
\end{tabular}
\begin{tablenotes}
\item \underline{Note}: The total numbers of intake gestures may slightly differ from Table \ref{tab:oreba-summary}. This is a technical implication of sampling with different frequencies (e.g., 8 fps for video), which can cause temporally close intake gestures to merge.
\end{tablenotes}
\end{threeparttable}
\end{table*}

Table \ref{tab:baseline} reports the test set results for the aforementioned models on both OREBA-DIS and OREBA-SHA.

\subsubsection{Inertial}

Heydarian et al. \cite{heydarian2020deep} ran multiple experiments benchmarking different deep learning models and pre-processing pipelines.
The top model performance was achieved by a CNN-LSTM with earliest fusion through a dedicated CNN layer and target matching.
Concerning preprocessing, their results show that applying a consecutive combination of mirroring, removing the gravity effect, and standardization was beneficial for model performance, while smoothing had adverse effects.

From the results in Table \ref{tab:baseline}, it appears that the test set for OREBA-DIS ($F_1=\fInertSOTADis$) is more challenging for inertial data than the test set for OREBA-SHA ($F_1=\fInertSOTASha$).
Comparing the simple CNN with the more advanced CNN-LSTM approach, we find that the more advanced CNN-LSTM add relative improvements of \relImpInertDis\% (OREBA-DIS) and \relImpInertSha\% (OREBA-SHA) over the simple CNN.

\subsubsection{Video}

Rouast and Adam \cite{rouast2019learning} applied several deep learning architectures established in the literature on video action recognition on the task of detecting intake gestures directly from the video data in OREBA-DIS.
The best test set result was achieved using a SlowFast \cite{feichtenhofer2018slowfast} network with ResNet-50 \cite{he2016deep} as backbone.
Further conclusions from the experiments are that appearance features are more useful than motion features, and that temporal context in form of multiple video frames is essential for top model performance.

The results in Table \ref{tab:baseline} indicate that the test set for OREBA-SHA ($F_1=\fVideoSOTASha$) is more challenging when working with video data than the test set for OREBA-DIS ($F_1=\fVideoSOTADis$).
Comparing results between the models, we find that the more advanced ResNet-50 SlowFast adds relative improvements of \relImpVideoDis\% (OREBA-DIS) and \relImpVideoSha\% (OREBA-SHA) over the simple CNN.

\section{Discussion}
\label{sec:discussion}

In this paper, we have introduced the OREBA dataset, which provides a comprehensive multi-sensor recording with labeled gestures of communal intake occasions from discrete and shared meals.
Building on a summary of related work on automatic dietary monitoring and an overview of existing public datasets in the field, we provided details on the data collection, sensor processing, and annotation methods employed in the creation of the OREBA dataset.
Additionally, we reported baseline results on the task of intake gesture detection based on video and inertial sensor data.

Sensor-based, passive methods of dietary monitoring have the potential of complementing existing active methods such as food records and 24-hr recall.
As seen in other fields such as object recognition \cite{deng2009imagenet} and action recognition \cite{kay2017kinetics}, progress in the research of machine learning methods is tightly linked to the availability and ongoing development of datasets with labeled examples.
In this light, we hope that the OREBA dataset will be able to support future developments in automatic dietary monitoring.
Compared to existing public datasets of labeled intake gestures, the OREBA dataset is unique as it is (i) multimodal with synchronised frontal video data based on spherical video recordings, and inertial data from both hands at 64 Hz, and (ii) includes a total amount of {\nPartTotal} recordings in two different communal eating scenarios.
As the first intake gesture detection dataset that also makes the video recordings available, OREBA enables researchers to independently verify, extend, and update the provided data annotations.
This further increases the transparency and reliability of the dataset, and the machine learning models building on it.

While we have reported results from our initial studies on detecting intake gestures with either video or inertial sensor data, there are also several other directions that research on this dataset could go into.
Sensor fusion of video and IMU for detecting intake gestures could be explored to combine the strengths of both approaches.
Further, the OREBA dataset can be used to compare and contrast CNN-LSTM \cite{kyritsis2019modeling} \cite{rouast2019learning} and CNN-GRU \cite{zhu2018sequence} models.
Improvements could also be made by using transfer learning between the two different scenarios, allowing to contrast the difficulties of monitoring discrete versus shared dishes.
Thanks to the availability of inertial data for both hands, studies could also explore how much information is retained on the dominant versus the non-dominant hand, which has implications for automatic dietary monitoring using commercial smartwatches.
Further, while our initial studies focused on detecting individual intake gestures, future studies could explore the other label categories -- for example, how well the video or inertial modalities perform at distinguishing between different utensils.
Finally, sensor fusion of video and IMU is a further possibility that could be explored in the future.
As such, and beyond researchers specifically interested in dietary monitoring, the OREBA dataset could also be a valuable resource for researchers interested in advancing machine learning models for sensor fusion more broadly.

\section{Conclusions}
\label{sec:conclusions}

Publicly available datasets are an important resource to fuel advances in machine learning \cite{deng2009imagenet} \cite{mollahosseini2017affectnet}.
In this paper, we have introduced a comprehensive multi-sensor recording dataset with labeled gestures of communal intake occasions from discrete and shared meals.
To the best of our knowledge, this is the first dataset for intake gesture detection that provides synchronized data in the form of both frontal video and inertial sensor data from both hands.
By making this dataset publicly available to the research community, OREBA has the potential to advance research and foster innovation in this area as it allows researchers to objectively benchmark existing and emerging machine learning approaches, and reducing the burden for researchers to collect and annotate their own data.

\section*{Acknowledgments}

We acknowledge the untiring support of Clare Cummings, Grace Manning, Alice Melton, Kaylee Slater, Felicity Steel and Sam Stewart in collecting and annotating the data.

\bibliographystyle{assets/IEEEtran}
\bibliography{paper}

\vskip -2\baselineskip plus -1fil

\begin{IEEEbiography}[{\includegraphics[width=1in,height=1.25in,clip,keepaspectratio]{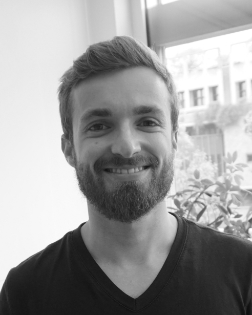}}]{Philipp V. Rouast}
received the B.Sc. and M.Sc. degrees in Industrial Engineering from Karlsruhe Institute of Technology, Germany, in 2013 and 2016 respectively.
He is currently working towards the Ph.D. degree in Information Systems and is a graduate research assistant at The University of Newcastle, Australia.
His research interests include deep learning, affective computing, HCI, and related applications of computer vision.
Find him at \url{https://www.rouast.com}.
\end{IEEEbiography}

\vskip -2\baselineskip plus -1fil

\begin{IEEEbiography}[{\includegraphics[width=1in,height=1.25in,clip,keepaspectratio]{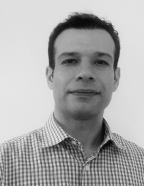}}]{Hamid Heydarian}
received the B.Sc. in Computer Engineering (software) from Kharazmi University, Iran, in 2002.
He is a senior software developer currently working towards the Ph.D. in Information Technology and is a casual academic at The University of Newcastle, Australia.
His research interests include inertial signal processing using deep learning and its related applications in dietary intake assessment and passive dietary monitoring.
\end{IEEEbiography}

\vskip -2\baselineskip plus -1fil

\begin{IEEEbiography}[{\includegraphics[width=1in,height=1.25in,clip,keepaspectratio]{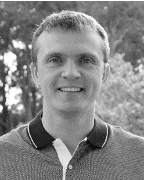}}]{Marc T. P. Adam}
is an Associate Professor in Computing and Information Technology at the University of Newcastle, Australia.
In his research, he investigates the interplay of human users' cognition and affect in human-computer interaction.
He is a founding member of the Society for NeuroIS.
He received an undergraduate degree in Computer Science from the University of Applied Sciences W{\"u}rzburg, Germany, and a PhD in Economics of Information Systems from Karlsruhe Institute of Technology, Germany.
\end{IEEEbiography}

\vskip -2\baselineskip plus -1fil

\begin{IEEEbiography}[{\includegraphics[width=1in,height=1.25in,clip,keepaspectratio]{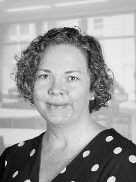}}]{Megan E. Rollo}
is a Research Fellow in Nutrition and Dietetics within the School of Health Sciences and Priority Research Centre for Physical Activity and Nutrition at The University of Newcastle, Australia.
She has received BAppSci, BHlthhSci(Nutr\&Diet), and PhD degrees from the Queensland University of Technology, Australia.
She has research interests in technology-assisted dietary assessment and personalized behavioral nutrition interventions.
\end{IEEEbiography}

\EOD
\end{document}